\documentclass[preprint,12pt]{elsarticle}

\usepackage{amssymb}
\usepackage{amsmath}
\usepackage[
linesnumbered,ruled,vlined]{algorithm2e}
\usepackage {algpseudocode}
\usepackage{algorithmicx}
\usepackage{algcompatible}
\usepackage{seqsplit}
\usepackage{hyperref}

\journal{Current Applied Physics}

\begin{document}

\begin{frontmatter}
\title{Machine learning-enhanced optical tweezers  for defect-free rearrangement}

\author{Yongwoong Lee}
\author{Eunmi Chae\corref{cor1}}
\ead{echae@korea.ac.kr}
\cortext[cor1]{Corresponding author}

\address{Department of Physics, Korea University, 145 Anam-ro, Seongbuk-gu, Seoul, 02841, Republic of Korea}

\begin{abstract}
Optical tweezers constitute pivotal tools in Atomic, Molecular, and Optical(AMO) physics, facilitating precise trapping and manipulation of individual atoms and molecules. This process affords the capability to generate desired geometries in both one-dimensional and two-dimensional spaces, while also enabling real-time reconfiguration of atoms. Due to stochastic defects in these tweezers, which cause catastrophic performance degradation especially in quantum computations, it is essential to rearrange the tweezers quickly and accurately. Our study introduces a machine learning approach that uses the Proximal Policy Optimization model to optimize this rearrangement process. This method focuses on efficiently solving the shortest path problem, ensuring the formation of defect-free tweezer arrays. By implementing machine learning, we can calculate optimal motion paths under various conditions, resulting in promising results in model learning. This advancement presents new opportunities in tweezer array rearrangement, potentially boosting the efficiency and precision of quantum computing research.
\end{abstract}





\end{frontmatter}


\section{Introduction}
Optical tweezer, a technique utilizing laser beams to trap and manipulate atoms or molecules, was proposed by Askin \cite{RN1,RN2}. When laser beams are applied to particles, they move towards the trapping point of the laser beams \cite{RN5, RN6, RN7}. Optical tweezers are well-suited for precise and delicate tasks to arrange locations of the trapped particles, finding applications in various research and applied fields. Particularly, in the field of Atomic Molecular and Optical(AMO) physics, optical tweezers have demonstrated remarkable achievements \cite{RN12, RN13, RN17}.

Atoms and molecules have been trapped and cooled in the optical tweezers starting from magneto-optical traps (MOTs) of them \cite{RN9, RN10, RN11}. One-dimensional as well as two-dimensional arrays of optical tweezers have beem implemented using Acouto-Optical Deflectors(AODs)\cite{RN12} and Spatial Light Modulators(SLMs) \cite{RN13, RN17, RN14, RN15, RN16}. These optical tweezer array capturing atoms and molecules have played a foundational role in advancing the technology underlying quantum computing and simulations \cite{RN18, RN63, RN64, RN66, RN67, RN68, RN69, RN70, RN71, RN72, RN73, RN74, RN75, RN76, RN77, RN78, RN79, RN80}.

However, when trapping atoms or molecules cooled from a MOT into an optical tweezer array, defect spaces arise due to the probabilistic nature of optical tweezer trapping \cite{RN23, RN24}. Therefore, it is necessary to first transform the optically tweezed array with defects into a defect-free state through optical tweezer rearrangement. One of the ways to rearrange the optical tweezers involves altering the Radio-Frequency(RF) signal's frequency components applied to AOD, moving the position of the optical tweezer array away from defect sites \cite{RN12}. Given the limited lifetime of optically trapped atoms, optical tweezer rearrangement must be completed with minimal movement in the shortest time possible \cite{RN26}. The inherent nature of optical tweezer rearrangement, resembling the shortest path problem, has led to the utilization of various algorithms such as the A* algorithm. Also recently, multiple algorithms for 2D optical tweezer rearrangement have been proposed \cite{RN29, RN30, RN31, RN32}. However, as optical tweezer rearrangement lacks an explicitly optimized path, the rearrangement time depends on both the defect sites within the optical tweezer array and the algorithm used \cite{RN30, RN31}. To address these challenges, we propose the introduction of reinforcement learning.

Reinforcement learning, a subset of machine learning, offers a methodology for the agent to learn how to maximize rewards for actions in an environment where the optimal actions are initially unknown \cite{RN33, RN34, RN35}. To apply reinforcement learning to solving problems related to a specific environment, understanding its states, actions, and rewards is essential \cite{RN34}. In reinforcement learning, an agent, using a neural network, interacts with the environment, and based on the changing state of the environment due to the agent's actions, obtains rewards for specific states \cite{RN35}. These rewards guide the reinforcement or inhibition of the agent's actions, helping the agent learn the non-linearities of the problem to be solved. The values at each step act as tuning parameters for the weights and biases of the neural network \cite{RN36}. The state of the environment is provided as input to the neural network, which outputs actions. Once a specific action value is output, the agent performs the corresponding pre-determined action, and this action changes the environment. Based on the altered environment, the agent receives rewards, and using these rewards, the neural network is tuned. Through this process, reinforcement learning enables the agent to learn optimal actions in complex environments.

One of the representative optimization problems is the shortest path problem \cite{RN37}. Reinforcement learning is suitable for solving the shortest path problem for various reasons. The shortest path problem typically involves finding the shortest route from a starting point to a destination in a complex environment. Reinforcement learning learns by interacting with the environment and accumulating experiences through trial and error. The agent decides which actions to take at each step, learns from the obtained rewards, and gains useful experiences for finding the shortest path \cite{RN38, RN39}. The shortest path problem requires considering future rewards, and reinforcement learning learns how to make decisions at each step to maximize future rewards, making it suitable for the shortest path problem. Moreover, the shortest path problem requires a balance between exploration and exploitation. The agent must utilize known paths while exploring new paths. Reinforcement learning learns and maintains this balance, helping to find optimal paths \cite{RN38}. Additionally, reinforcement learning algorithms can be applied to various types of optimization problems \cite{RN40}.

The optical tweezer rearrangement problem can be transformed into a shortest path problem. Therefore, we used reinforcement learning to solve the problem of rearranging optical tweezers in a better way. This method helps us quickly find the right steps to rearrange the tweezers, especially when they are used to trap atoms or molecules. This is important because the way the tweezers need to be rearranged changes depending on the defects in the system, and reinforcement learning helps us figure out the best rearrangement for each situation.

\section{Methodology}
To handle the rearrangement of optical tweezers through reinforcement learning, it is essential to first define the agent and environment. To define these, we need to examine the setup environment for optical tweezers. For the simulation environment for optical tweezer rearrangement, we assumed using an AOD to generate the optical tweezer array. An AOD is a device that uses RF signals to change the direction of laser beams \cite{RN41, RN42}. It generates high-frequency sound waves within a crystal or similar material. When laser beams pass through, they interact with these sound waves. This interaction causes the laser beam's direction to deflect, allowing precise control of the direction by adjusting the frequency of the sound waves. If an RF signal with multiple frequencies, i.e., 
\begin{eqnarray}
    \mathrm{RF\,signal} = \sum_{i=1}^{n}A_i \mathrm{sin}(2\pi f_i t)
\end{eqnarray}\label{eqn:1}is applied to the AOD, the laser will be split into \textit{n}, the number of multitone frequencies in the RF signal, under the resolution of the AOD. Therefore, when independent frequencies are synthesized and applied to the AOD, the incident laser in the crystal's reaction range of the AOD will split into the number of frequencies applied.

When particles are stochastically captured in the tweezer array, empty spaces, i.e., defect sites, are created \cite{RN9, RN12, RN14}. The presence or absence of particles at each tweezer location is observed using the fluorescence of the particles and detected by a CCD camera. To transform an array with defects into a defect-free state, it is necessary to change the frequency components corresponding to the optical tweezers where particles are trapped. As each tweezer corresponds to a specific frequency, to move the optical tweezer array, the frequency of the optical tweezer where particles are trapped, needs to be changed.

If the frequency of the defect optical tweezer site corresponds to $f_m$ and the frequency of the optical tweezer site where an atom is trapped corresponds to $f_n$, to creat a loop for generating an RF signal, we repeat the loop until 
\begin{eqnarray}
    f_m = f_n + \Delta f \cdot \mathrm{iteration}
\end{eqnarray}\label{eqn:2}is satisfied. Here, the value $\Delta f$, change of the frequency per step, is a variable that depends on experimental setup.

\begin{figure}[t]
	\centering
	\includegraphics[width=90mm]{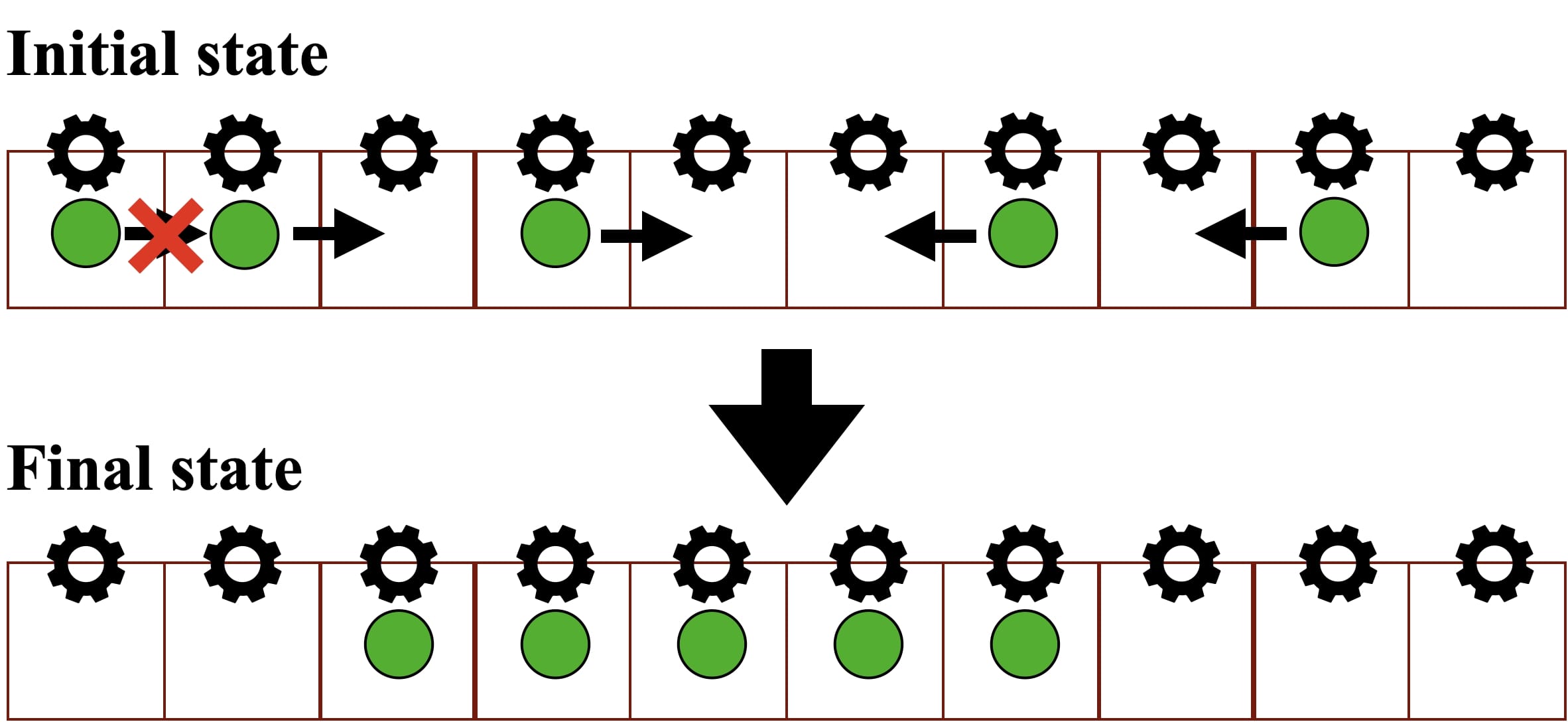}
	\caption{A schematic representation of the rearrangement of the tweezers. Each machine(box with the cog) is aligned side side and a green sphere in a box depicts a trapped atom in the optical tweezer. The direction of the arrow indicates the direction in which the atom can move, and a red prohibition marker means that the action is prohibited when another atom lies in the direction of the atom's movement. The terminate state is the state in which the atoms are arranged to a defect-free state in the center of the array.
    }
	\label{FIG:1} 
\end{figure}

Ultimately, continuously iterating the loop until $f_m = f_n$ for all $f_m$ values will result in obtaining the defect-free state. For the simulation, we can simplify the operation to set $\Delta f$ to the interval of each tweezers. We can envision this kind of machine as shown in Figure \ref{FIG:1}. This machine is assumed to move atoms to the left or right one by one when there is an atom in the device, and stay still when there is none. However, the machine cannot move the atom if the neighbor sites are occupied by atoms. With such machines arranged in a row, all atoms will eventually converge towards the center of the array.

To summarize, the mechanical action of moving an object sideways equals to change the frequency value. The presence of objects in the machine corresponds to a situation where particles are trapped in an optical tweezer, and the absence of objects in the machine corresponds to a situation where particles are not captured in the optical tweezer, i.e., a defect site. Therefore, if we simulate this machine we can simulate a tweezer.

Setting up a simulation environment for reinforcement learning, we need to mimic a real-world lab environment. Generally, when atoms are trapped in an optical tweezer, the trapping probability is approximately $p\sim0.6$ \cite{RN10,RN12}. The number of optical tweezers for the simulation ranged from 50 to 300, with 10 increments each time from 50 to 100, followed by 200 and 300. The termination conditions are set either when atoms fill 40\% of the total arrays at the center region, or when the time step within the episode exceeds 10,000. The definition of states, actions, and rewards for the reinforcement learning problem are set as follows:

\begin{figure}[t]
    \centering
	\includegraphics[width=90mm]{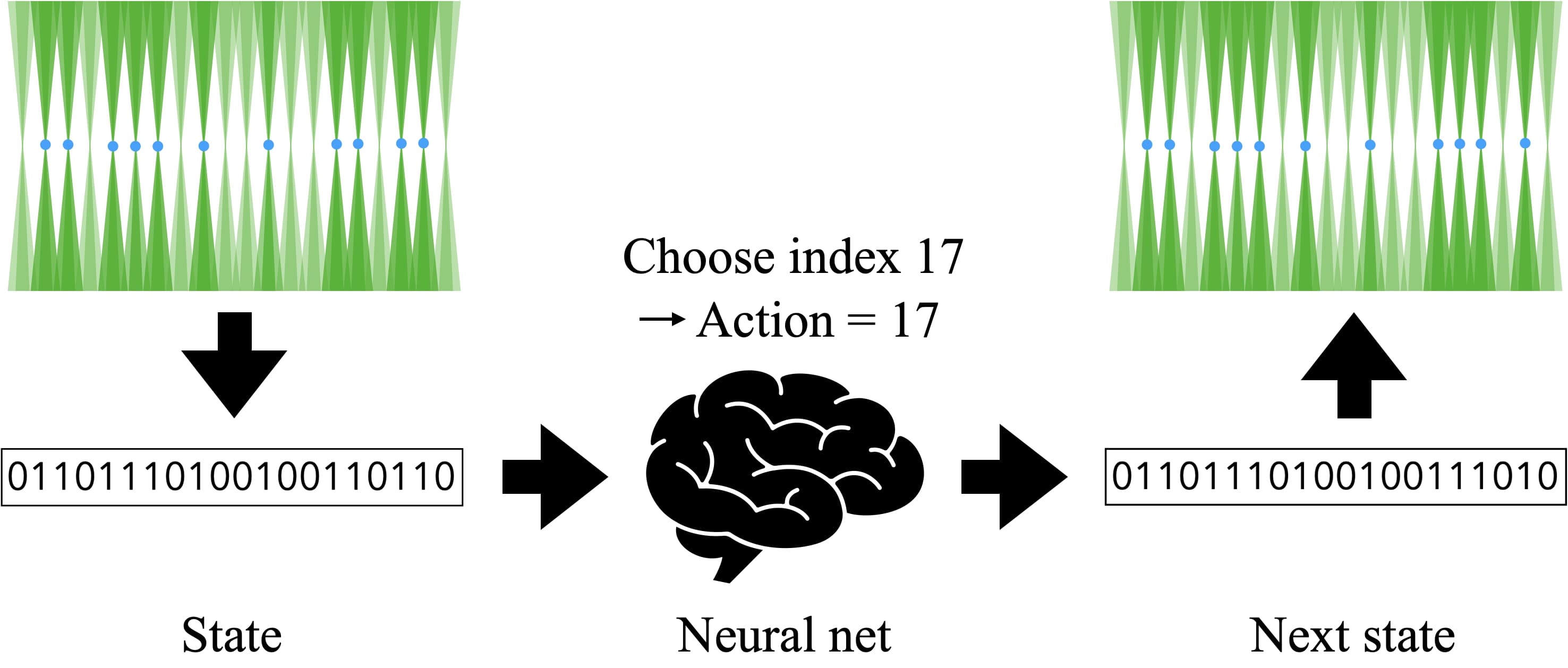}
	\caption{A schematic of how neural network works. The array contains randomly loaded atoms, which can be classified as either defective (shown in light green) or not defective (shown in dark green). This classification creates a binary array of states, where defect sites are represented by 0 and sites with atoms are represented by 1. The neural network then computes the state value and generates an action value, which acts as an index value for the binary array. The transition to the next state occurs according to a preset action algorithm.
    }
	\label{FIG:2}
\end{figure}

\begin{flushleft}
\textbf{\textbullet \,State}
\end{flushleft}
If we express tweezer sites where atoms are captured as 1 and sites where atoms are not captured as 0, we obtain a binary array of the same length as the number of optical tweezer array. The probability of capturing atoms in each tweezer is independent, and the average capture probability is $p\sim0.6$. Thus, the initial states of the optical tweezer arrays form a Bernoulli distribution with a probability of 0.6. To further use information from the 40\% center area, the position of the final defect-free array, as a state function, the binary values from that region are extracted separately, creating another binary array. The remaining positions outside this region are marked as 0, creating a binary array of the same size as the one generated from the Bernoulli distribution. These two binary arrays are combined into one batch and processed as the input value of a 1-dimensional Convolutional Neural Network(CNN) \cite{RN43}.

\begin{flushleft}
\textbf{\textbullet \,Action}
\end{flushleft}
 We define the action as the act of determining which machine to operate. Therefore, the maximum size of action space is the same as the number of machines. The direction of the atom's movement is determined by its position relative to the center of the entire array. This reduces the degree of freedom of the atomic movement and eases the learning process. Also, the machine only moves the atom if it contains an atom. Therefore, each action follows \textit{Algorithm 1}.

\begin{algorithm}[!t]
\caption{Action algorithm}
\textbf{Input:} state, total\_array\_size, action

\textbf{Output:} Next\_State

\medskip
Define \textbf{State} as \textit{s}

Define \textbf{Next\_State} as \textit{$s'$}

Define \textbf{Total\_Array\_Size} as \textbf{$T_{\mathrm{size}}$}

Define \textbf{Is\_Filled} as \textit{s}[\textit{a}]

Define \textbf{Is\_Empty\_R} as \textit{s}[\textit{a+1}]

Define \textbf{Is\_Empty\_L} as \textit{s}[\textit{a-1}]

Choose action \textit{a}

\If{Is\_Filled = 1}{
    \If{$a < T_{\mathrm{size}}//2$}{
        \If{Is\_Empty\_R = 0}{
        $s[a], s[a+1] = 0, 1$
        }
        \Else{
        pass
        }
    }
    \Else{
        \If{Is\_Empty\_L = 0}{
        $s[a-1], s[a] = 1, 0$
        }
        \Else{
        pass
        }
    }
}
\Else{
pass
}

\textit{$s'$} = \textit{s}

\Return \textit{$s'$}
\end{algorithm}

\begin{flushleft}
\textbf{\textbullet \,Reward}
\end{flushleft}
The reward function is a factor in reinforcement learning that directly influences the updates of the neural network during learning. To ensure that learning proceeds to the desired direction, the reward function needs to be appropriately set. Positive rewards and negative rewards are used to achieve this:
\\\\
\quad\text{\textbullet \,Positive reward}
\\\\
The number of atoms captured in the tweezer array at the region of interest is assessed. If atoms are located in the region of interest, each atom gains a positive reward $r=0.1$. When the final state is reached, a reward of $r=1$ is given.
\\\\
\quad\text{\textbullet \,Negative reward}
\\\\
Negative rewards are given for actions that do not lead to meaningful movements. If a machine is selected, but does not contain an atom or if, even with an atom, the neighboring machine contains an object inside it, a reward of $r=-0.1$ is given for such actions. Additionally, to reach the final state as quickly as possible, a reward of $r=-0.01$ is given at each time step.

\begin{flushleft}
\textbf{\textbullet \,Termination state}
\end{flushleft}
The termination state is defined as a situation when the region of interest is filled with atoms without any defects, or when the episode reaches the maximum time step. Tweezer sites with atoms locate outside of the region of interest are independent of the termination state. Therefore, their positions are random outside.

A visual representation of each of the state, action defined above, and the operating sequence are depicted in Figure \ref{FIG:2}. 

\section{Training method}

\begin{figure}[htbp]
    \centering
    \includegraphics[width=90mm]{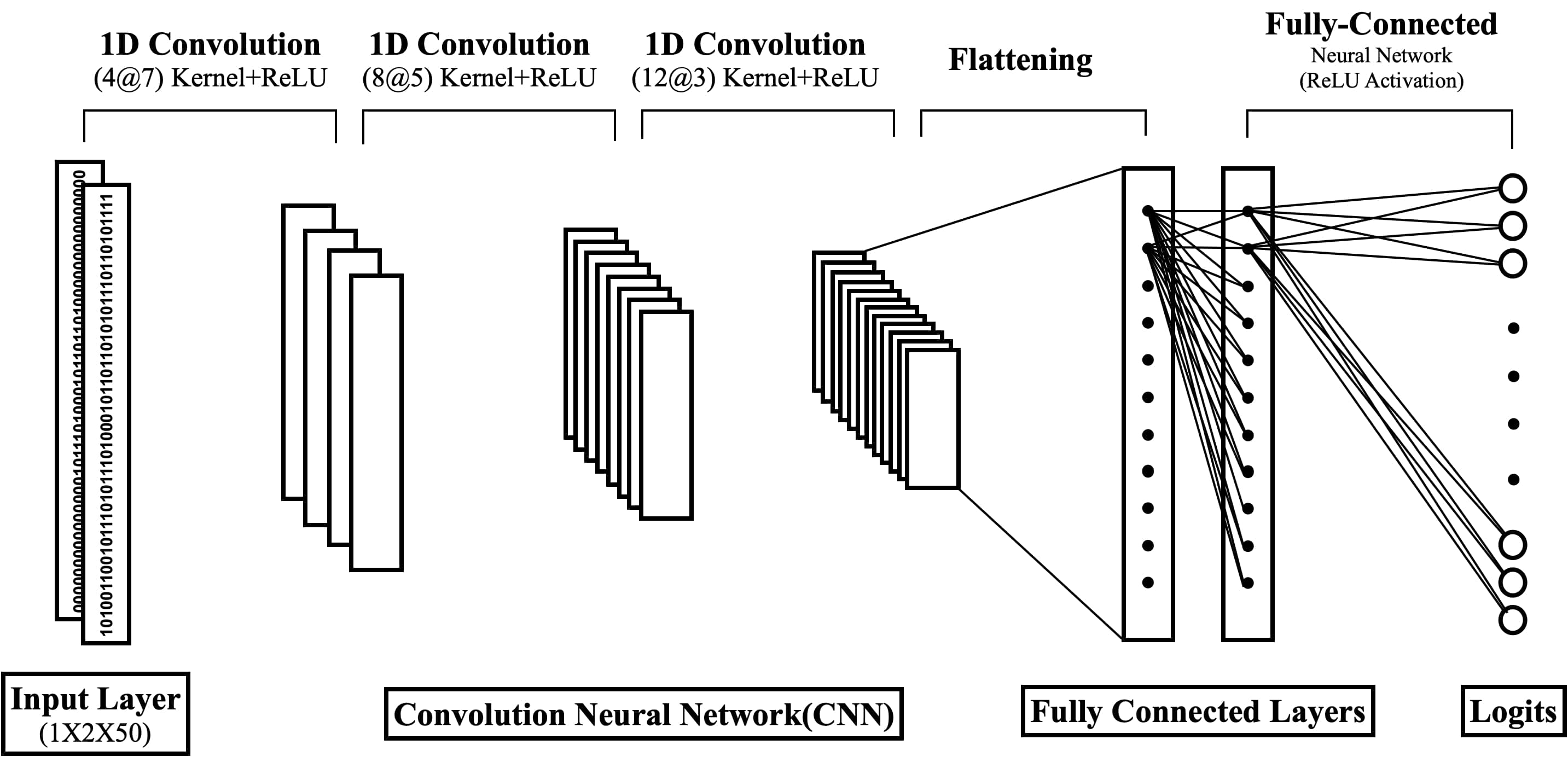}
    \caption{Example of the actor network structure when the total optical tweezer array size is 50.}
    \label{FIG:4}
\end{figure}

We used the Proximal Policy Optimization(PPO) algorithm \cite{RN44} as a reinforcement learning algorithm to solve our problem. The PPO algorithm is an on-policy algorithm, which is applicable in environments with continuous action spaces and discrete action spaces, and is especially effective in large action spaces \cite{RN44,RN45,RN46}.
\begin{figure}[htbp]
    \centering
    \includegraphics[width=30mm]{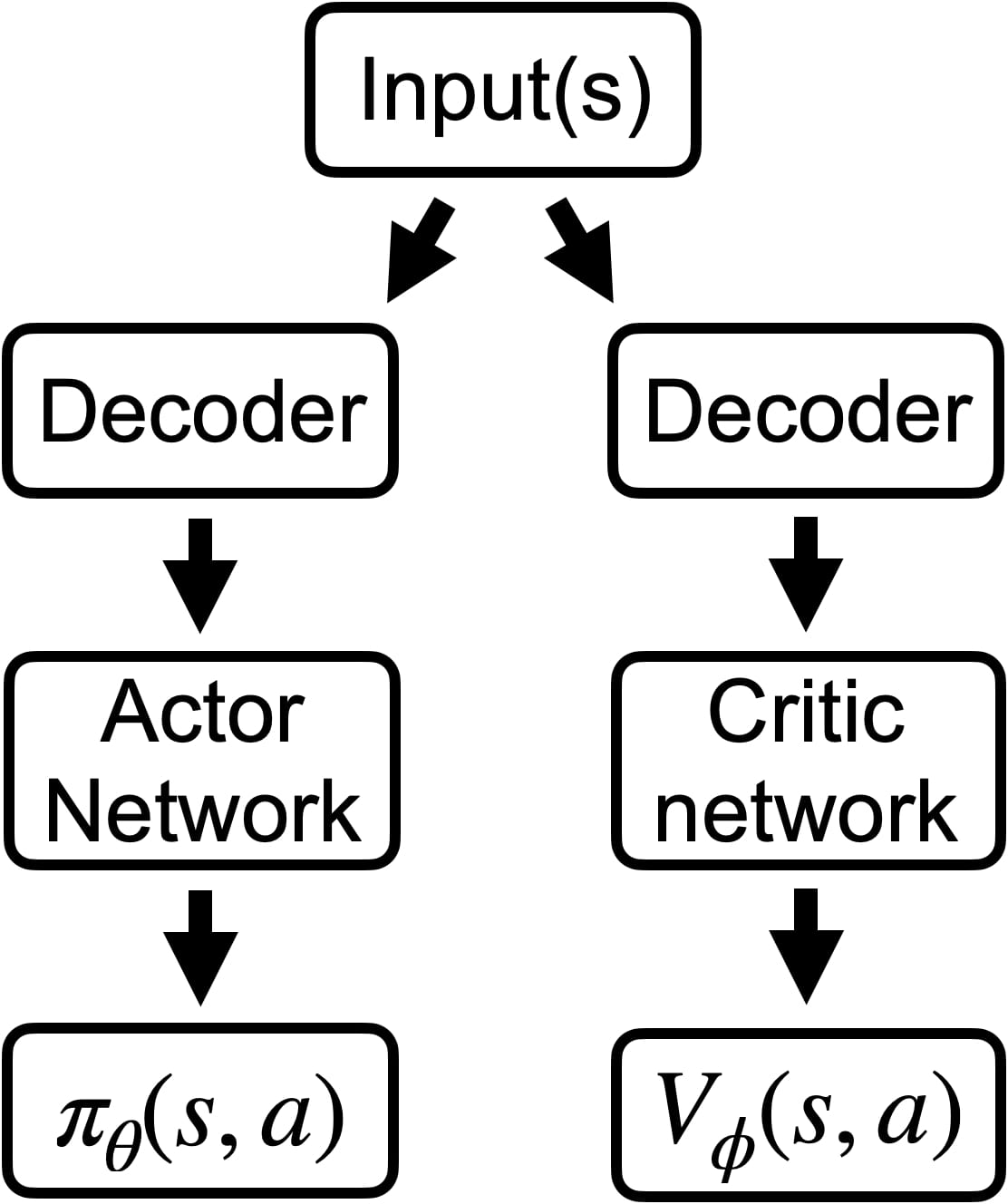}
    \caption{A schematic of decoupled actor-critic network structure.}
    \label{FIG:3}
\end{figure}
Since the action space for the optical tweezer array is composed of a large, discrete action space from 50 to 300, we adopted the PPO algorithm to solve the reinforcement learning problem. The structure of the entire neural network is composed of a decoupled actor-critic network(Figure \ref{FIG:3}) \cite{RN47,RN48,RN49} to achieve fine-tuning for each network. The overall structure of the neural network is shown in Figure \ref{FIG:4}. 
\begin{algorithm}[t]
\caption{PPO training algorithm}
\textbf{Input:} initial policy parameter $\theta$, initial value function parameter $\phi$

\medskip
Initialize actor network parameter \textbf{$\theta$}

Initialize critic network parameter \textbf{$\phi$}

\For{$0 \leq i \leq K_{\mathrm{epoch}}$}{
    Gether $\tau_{i} = \{s_{i}, a_{i},r_{i}, s'_{i}\}$
    
    Compute ${A}^{\mathrm{GAE}}_{\theta_{t}}$
    
    $\theta_{i+1} = \underset{\theta}{\mathrm{argmax}}\frac{1}{|\tau_i|K_{\mathrm{epoch}}}\sum\limits_{t=0}^N L_{\theta_t}$

    $\phi_{i+1} = \underset{\phi}{\mathrm{argmin}}\frac{1}{|\tau_i|K_{\mathrm{epoch}}}\sum\limits_{t=0}^N L_{\phi_t}$

    Soft update parameters via Adam optimizer
    
    $\theta'\leftarrow \tau\theta + (1-\tau)\theta'$

    $\phi'\leftarrow \tau\phi + (1-\tau)\theta'$
}
\textbf{end for}
\end{algorithm}
The overall structure of the neural network, which is common for both actor and critic networks, is shown in Figure \ref{FIG:4}. For the state function \textit{s}, we created two batches of the binary array: one full binary array representing the information of the occupancy in the optical tweezer array, and one binary array that only contains the occupancy information of the region of interest. The size of each binary array is equal to the size $T_{\mathrm{size}}$ of the total tweezer array, so the dimension of the state function is defined as $1\cdot2\cdot T_{\mathrm{size}}$. The state function is processed through the CNN filters of the actor network and the critic network. The data with the dimension of $1\cdot2\cdot T_{\mathrm{size}}$ is refined by batch normalization and rectified linear unit activation function(ReLU)\cite{RN50} while passing through the CNN layer where the kernel size is reduced in the order of 7, 5, 3 and the number of filters is increased in the order of 4, 8, and 12. The final dimension of the encoded data is $1\cdot12\cdot(T_{\mathrm{size}}-12)$. 

After passing through the CNN layer, the state information encoded in the $1\cdot12\cdot(T_{\mathrm{size}}-12)$ dimension is flattened and set as the input to the Fully-Connected(FC) layer. When the total tweezer array size is 50 to 100, flattened data is connected to the FC layer with a size of 128, and if it is 200 and 300, it is connected to the FC layer with a size of 512. The outputs of the FC layer are \emph{$\pi_{\theta}(s ,a)$} in $1\cdot T_{\mathrm{size}}$ dimension for the Actor network and \emph{$V_{\phi}(s)$} in $1\cdot1$ dimension for the critic network. The output dimension size and number of parameters for each neural network for $T_{\mathrm{size}}$ can be found in Table \ref{tab:1}. The information used to update the neural network are the initial state \textit{s} corresponding to the input data of the neural network, the output data $\pi_{\theta}(s,a)$ of the actor network, and a probability distribution generated by softmax using $\pi_{\theta}(s,a)$. Here, the index of the most probable value is taken as the action \textit{a}. Action \textit{a} causes a state transition, and the environment provides information about the next state, \emph{$s'$}, and the reward \emph{$r$} for the action. \emph{$s'$} again becomes the input data of the neural network, and the loop continues until $s'$ reaches a terminating state or the maximum time step is reached. At the end of each loop, the information about \emph{$s, s', \pi_{\theta}(s, a), a, r$} is transferred to a mini-batch, and a batch of size \textit{N} is created for training every \textit{N} time step. The information contained in this mini-batch is used to update the neural network by the PPO algorithm. 
\begin{table}[t]
\centering
\scriptsize{
\caption{Total number of neural network parameters}
\begin{tabular}{ccc}
\hline
\textbf{Tweezer array size($T_{\mathrm{size}}$)}                            & \multicolumn{1}{c}{\textbf{50-100}} & \multicolumn{1}{c}{\textbf{200-300}} \\ \hline
\multicolumn{1}{c}{\textbf{Layer(type)}} & \multicolumn{2}{c}{\textbf{Number of parameters}}                                      \\ \hline
\multicolumn{1}{c}{Conv1d-1}             & \multicolumn{1}{c}{$60$}              & \multicolumn{1}{c}{$60$}                   \\ 
\multicolumn{1}{c}{BatchNorm1d-2}        & \multicolumn{1}{c}{8}               & \multicolumn{1}{c}{8}                      \\ 
\multicolumn{1}{c}{Conv1d-3}             & \multicolumn{1}{c}{168}             & \multicolumn{1}{c}{168}                   \\ 
\multicolumn{1}{c}{BatchNorm1d-4}        & \multicolumn{1}{c}{$16$}              & \multicolumn{1}{c}{$16$}                   \\ 
\multicolumn{1}{c}{Conv1d-5}             & \multicolumn{1}{c}{$300$}             & \multicolumn{1}{c}{$300$}                 \\ 
\multicolumn{1}{c}{BatchNorm1d-6}        & \multicolumn{1}{c}{$24$}              & \multicolumn{1}{c}{$24$}                     \\ 
\multicolumn{1}{c}{Linear-7} &
  \multicolumn{1}{c}{$128\cdot12\cdot(T_{\mathrm{size}}-12) + 128$} &
  \multicolumn{1}{c}{$512\cdot12\cdot(T_{\mathrm{size}}-12) + 512$} \\ 
\multicolumn{1}{c}{Linear-8} &
  \multicolumn{1}{c}{$128\cdot T_{\mathrm{size}}+T_{\mathrm{size}}$} &
  \multicolumn{1}{c}{$512\cdot T_{\mathrm{size}}+T_{\mathrm{size}}$} \\ \hline
\end{tabular}
\label{tab:1}}
\end{table}

\begin{table}[t]
    \centering
    \scriptsize{
    \setlength{\tabcolsep}{0.3\tabcolsep}
    \caption{Hyperparameters for training}
    \begin{tabular}{ccccc}
    \hline
    \multicolumn{2}{c}{\textbf{Total tweezer array size}} & \multicolumn{1}{c}   {\textbf{50-100}} & \multicolumn{1}{c}{\textbf{200}} & \textbf{300} \\ \hline
    \multicolumn{1}{c}{\textbf{Hyperparameter}} & \textbf{Meaning}                        & \multicolumn{3}{c}{\textbf{Value}}                                \\ \hline
    \multicolumn{1}{c}{\textit{N}}                       & Batch size                              & \multicolumn{1}{c}{$32$}     & \multicolumn{1}{c}{$128$}    & $512$   \\ 
    \multicolumn{1}{c}{$K_{\mathrm{epoch}}$}               & Number of iteration & \multicolumn{1}{c}{$8$}      & \multicolumn{1}{c}{$8$}      & $8$     \\ 
    \multicolumn{1}{c}{$\gamma$}                  & Discount factor                         & \multicolumn{1}{c}{$0.98$}   & \multicolumn{1}{c}{$0.98$}   & $0.98$   \\ 
    \multicolumn{1}{c}{$\lambda$}                 & Bias-Variance tradeoff factor           & \multicolumn{1}{c}{$0.95$}   & \multicolumn{1}{c}{$0.90$}   & $0.90$   \\ 
    \multicolumn{1}{c}{$\tau$}                    & Learning rate                           & \multicolumn{1}{c}{$2\cdot10^{-4}$} & \multicolumn{1}{c}{$2\cdot10^{-4}$} &$2\cdot10^{-4}$ \\ \hline
    \end{tabular}
    \label{tab:2}}
\end{table}
Updating the parameters of the actor network and the critic network requires several operations with the data in the batch. First, we need to find the Temporal Difference(TD) target and TD error($\delta_{t}$). TD is mainly used to train the value function using the actual reward and the future estimated value of the next step, but here, it is used to train both the policy function and the value function. The TD target and $\delta_{t}$ are calculated as follows.
\begin{align}
    &\mathrm{TD}_{\mathrm{target}} = r + \gamma V_{\phi}(s')\label{eqn:4}
    \\
    &\delta_{t} =  r + \gamma V_{\phi}(s') - V(s;\phi)\label{eqn:5}
\end{align}
The $\gamma$ here is called the discount factor, a hyperparameter used to convert future value to present value. With the value obtained in equation (\ref{eqn:5}), we can estimate the advantage $A(s, a;\phi)$ using Generalized Advantage Estimation(GAE)\cite{RN51}, a method for finding the advantage $A(s, a;\phi)$ that needs to be obtained to update the actor network.
\begin{eqnarray}
    A_{\phi}^{\mathrm{GAE}}(s_{t}, a_{t}) = \sum\limits_{i=t}^{t+N-1}(\gamma\lambda)^{i-t}\delta_{i}\label{eqn:6}
\end{eqnarray}
Using equation (\ref{eqn:6}), the advantage is estimated from the start of the input batch data, \textit{t}, through the batch size, \textit{N}, of data. The value of $\lambda$ is a term that trades off the variance and bias of the advantage estimate to get an appropriate value. For a stable learning, we now use a method called clipping that constrains the ratio of the current policy function $\pi_{\theta}(s, a)$ to the previous policy function $\pi_{\theta_{\mathrm{old}}}(s, a)$. The value of $\epsilon$ is defined as keeping the ratio between $1-\epsilon$ and $1+\epsilon$. The objective function obtained using this method is shown below.
\begin{multline}
    L_{\theta} = min\Big(\frac{\pi_{\theta}(s,a)}{\pi_{\theta_{old}}(s,a)} A^{\pi_{\theta_{old}}}(s,a),\\
    clip\Big(\frac{\pi_{\theta}(s,a)}{\pi_{\theta_{old}}(s,a)}, 1-\epsilon, 1+\epsilon \Big) A^{\pi_{\theta_{old}}}(s,a) \Big)
\label{eqn:7}
\end{multline}
The objective function for updating the critic network parameters is as follows
\begin{eqnarray}
    L_{\phi} = (r + \gamma V_{\phi}(s') - V_{\phi}(s))^2
\label{eqn:8}
\end{eqnarray}
With that, we are ready to update the parameters of the actor network and critic network with the mini batch information. The PPO pseudocode structure for updating parameters is shown in \textit{Algorithm 2}. The neural network was trained by iterating over up to 10,000 episodes, using up to 10,000 time steps for the total tweezer array size of 50 to 100, and up to 200,000 time steps for 200 and 300. The hyperparameter tuning values for training are as follows Table \ref{tab:2}.

\section{Training results}
To evaluate whether they had learned correctly, we give a score to the agent. The scoring system for the reinforcement learning model is based on the successful movement of atoms within the optical tweezer array. If the agent moves an atom to the right site in the array, where there is already an atom and no atoms nearby, it gets a score of 1. Otherwise, a score of 0 is assigned. The score for each episode is then divided by the total time steps in the episode to determine the accuracy of each episode's learning. Additionally, the number of time steps taken to reach the termination state for each episode is measured.

\begin{figure*}[t]
    \centering
    \includegraphics[width=140mm]{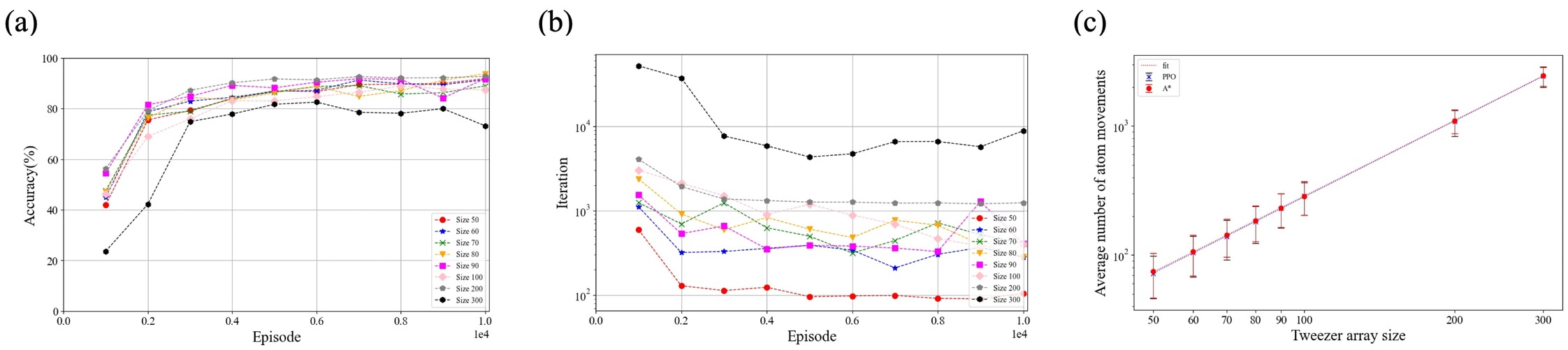}
    \label{FIG:5a}
    \label{FIG:5b}
    \label{FIG:5c}
    \caption {Analysis of training results of the simulation. The total tweezer array size ranges from 50 to 300. (a), (b) show an average accuracy and an average iteration of each array size for an episode with a total length of 10,000. (c) shows an average number of atom movements of each array size comparing two different algorithms using PPO and A* algorithms. Each axis is expressed in a logarithmic scale. Each iteration is an average of 1,000 simulations}
\end{figure*}

Figure \ref{FIG:5a} illustrates the accuracy for each episode based on various optical tweezer array sizes. Accuracy is calculated by dividing the score for each episode by the length of time step in each episode. It was observed that the accuracy increased for all array sizes throughout 10,000 episodes. The accuracy values converged to around 90\% except for the size 300 case, that the neural network has properly functioned acting as a classifier determining the atoms' locations and which atoms to move in a given state to reach the termination state.

Figure \ref{FIG:5b} depicts the episode-wise iteration count for each optical tweezer array size. Since we aimed for the optical tweezer array to reach the defect-free state in the shortest time possible, if the learning proceeded smoothly, the iteration count should decrease with each episode. Figure \ref{FIG:5b} confirms a trend of decreasing iteration count as the episodes progress. Additionally, as the total tweezer array size increases, there is a tendency for the iteration count to increase. This is a natural tendency because as the tweeter array size increases, the number of atoms that need to be moved increases. A common feature of Figures \ref{FIG:5a} and \ref{FIG:5b} is that size 300 deviates from the trend of the other learning results at around 10,000 episodes. This is a problem caused by overfitting. As the action space increases, the complexity of the loss function and the exploration space also increases. It leads to decreasing sample efficiency and potential overfitting, making it harder to achieve optimal policy convergence. This can be improved by more precise hyperparameter tuning. Also, even though overfitting occurred, there is a large trend of increasing score and decreasing iteration, so that we can conclude the learning was successful.
\begin{table}[b]
\centering
\caption{The average number of atom movements for each size of the tweezer array}
\begin{tabular}{ccc}
\hline
\textbf{Total tweezer array size} & \textbf{PPO algorithm} & \textbf{A* algorithm} \\ \hline
$50$                        & $72\pm26$              & $75\pm28$             \\ 
$60$                        & $104\pm37$             & $106\pm37$            \\ 
$70$                        & $139\pm47$             & $143\pm47$            \\ 
$80$                        & $181\pm58$             & $184\pm57$            \\ 
$90$                        & $230\pm68$             & $231\pm67$            \\ 
$100$                       & $287\pm84$             & $284\pm79$            \\ 
$200$                       & $1087\pm250$           & $1100\pm223$          \\ 
$300$                       & $2445\pm445$           & $2452\pm405$          \\ \hline
\end{tabular}
\label{tab:3}
\end{table}
\begin{figure}[t]
    \centering
    \includegraphics[width=90mm]{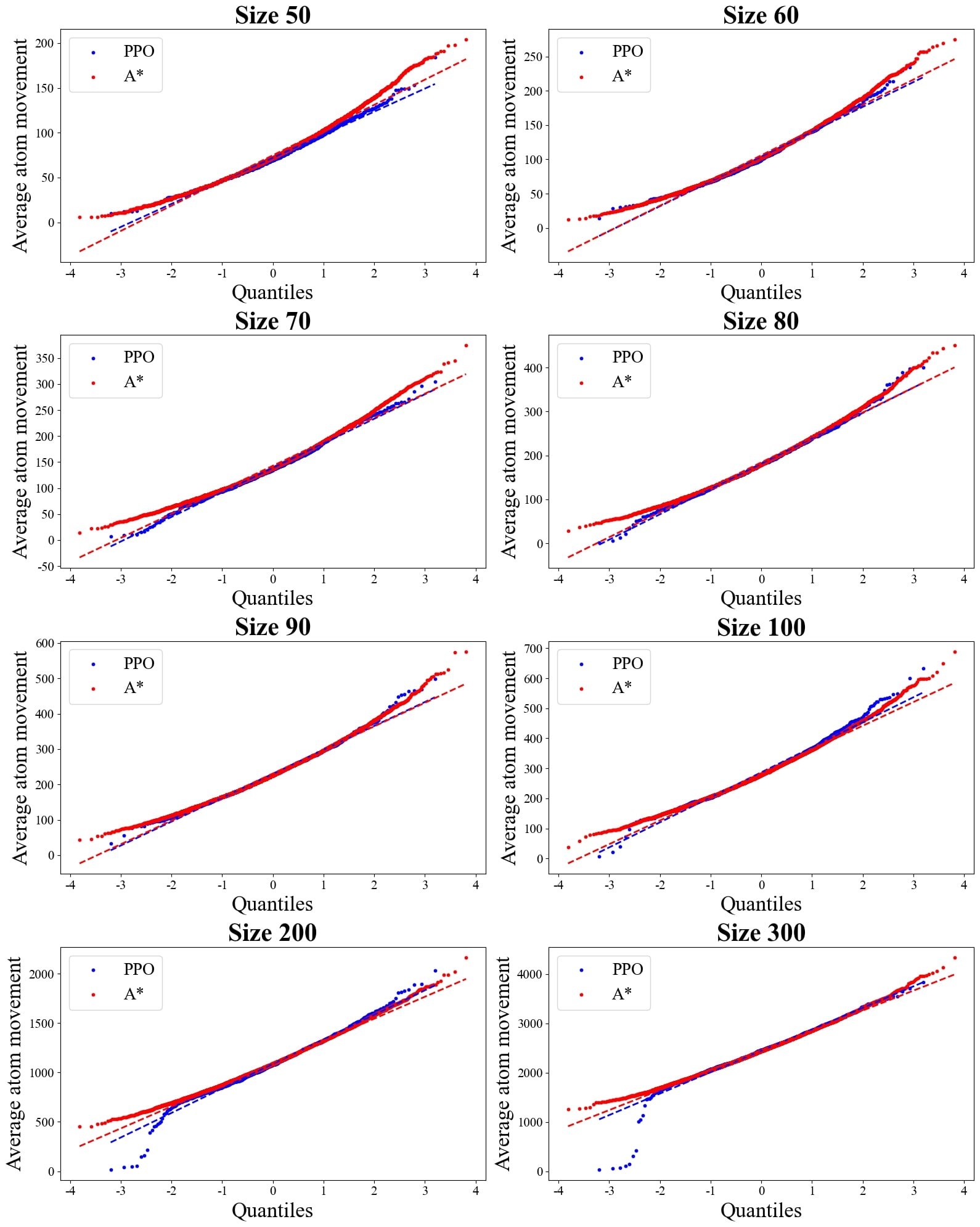}
    \caption{Q-Q plots of the average numbers of atom movements for array sizes from 50 to 300}
    \label{FIG:6}
\end{figure}
Figure \ref{FIG:5c} and Table \ref{tab:3} displays the average atom movement counts for a total of 1,000 episodes using a randomly generated binary array with $p\sim0.6$. Each data was given using the pre-learned neural network parameters for each optical tweezer array size. The A* algorithm was utilized to compare the performance of the reinforcement learning model. The A* algorithm acted as an agent moving atoms closest to the termination state for each episode, selecting actions at each time step to minimize the cost and reach the goal node \cite{RN52,RN54}. The average atom movement count obtained through the A* algorithm served as a baseline for evaluating the PPO algorithm pre-trained neural network model. Table 3 provides specific values for Figure \ref{FIG:5c}, indicating the average and standard deviation of the number of the average atomic movements for each algorithm across all optical tweezer array sizes. The results of the two algorithms are consistent, and both show a quadratic increase in the size of the tweezer array, as indicated by a fitting index of about 1.95. It can be concluded that the neural network trained through reinforcement learning functions well as a classifier capable of distinguishing which atoms are in a movable state and should move to reach the termination state in the shortest time.

Figure \ref{FIG:6} represents Quantile-Quantile plots (Q-Q plots) for 1,000 data points for each algorithm using the average movement counts obtained in Figure \ref{FIG:5c}. A Q-Q plot is a graphical tool used to check if two datasets follow the same distribution. It displays the quantiles of each dataset on the x-axis and the corresponding values of the data samples on the y-axis \cite{RN55}. If the Q-Q plot appears linear, it indicates that the distribution corresponds to a normal distribution. Using this graph, we could visually inspect the distribution of average atom movement counts obtained by both algorithms. Overall, a linear relationship was observed up to 2 sigma around the median of each algorithm's dataset. However, beyond this point, it appears that the linearity is lost, suggesting that each distribution loses normality for outliers. Another interesting aspect is the noticeable difference in the distribution of algorithmic data for parts where the number of average atom movement is small, especially as the overall array size increases. In the A* algorithm, the slope of the data distribution is gradual for the quantile region between -4 and -2. In contrast, the PPO algorithm shows a distribution where the values are clustered near 0 in this region.

The reason for this phenomenon is overfitting. When a specific state does not return appropriate action values for a successful state transition in an episode, the atom remains in the state and does not move. Then the count of the atom movement becomes near 0 when the maximum time step is reached. This problem can be addressed during training through the hyperparameter tuning.

\section{Training result in two-dimensional optical tweezer rearrangement}
Our developed reinforcement learning method is easily scalable in dimension. A two-dimensional (2D) array of optical tweezers is a direct extension of the one-dimensional(1D) case. It can be generated by employing two AODs placed along the x-axis and y-axis \cite{RN13,RN31}. When the light passes the x-axis AOD first followed by the y-axis AOD, each tweezer generated by the x-axis AOD branches again by the y-axis AOD, forming a 2D tweezer array. Therefore, the two axes of the array are dependent to each other, which makes the rearrangement challenging. To circumvent the issue, a total of four AODs are assumed in the simulation: two for static traps and the other two for mobile traps. 

\begin{figure*}[t]
    \centering
    \includegraphics[width=140mm]{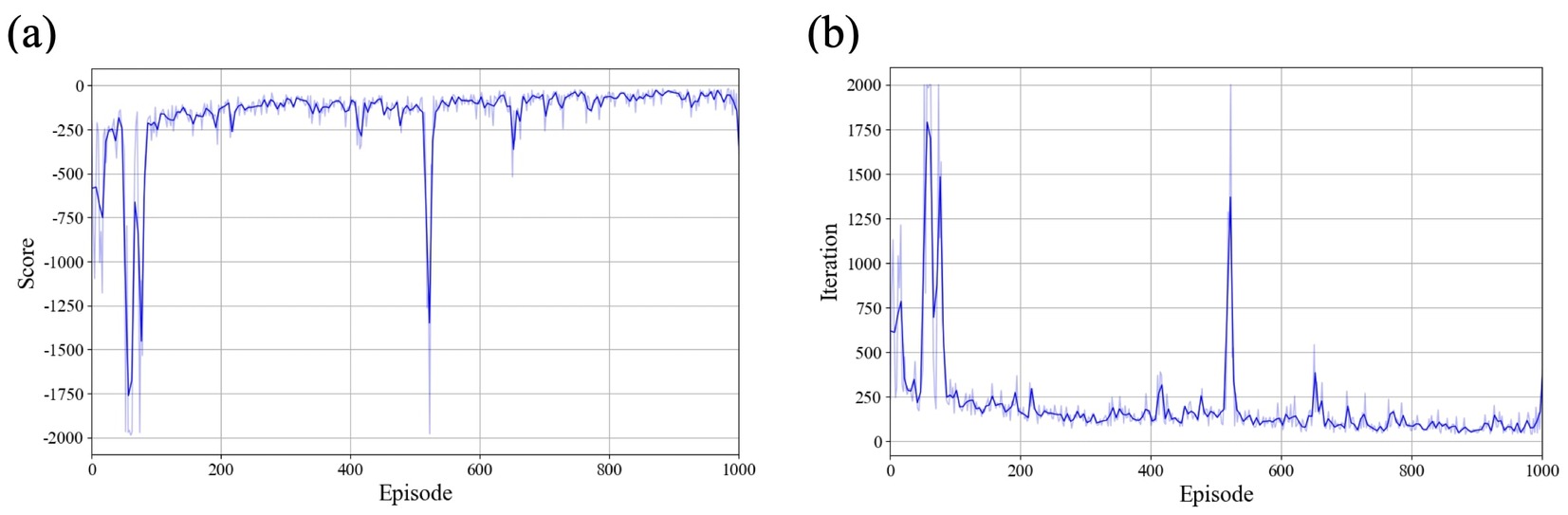}
    \caption{Training results of the rearrangement simulation for a 2D tweezer array. (a) and (b) show the score and iteration of each episode.}
    \label{FIG:7}
\end{figure*}

\begin{figure}[t]
    \centering
    \includegraphics[width=90mm]{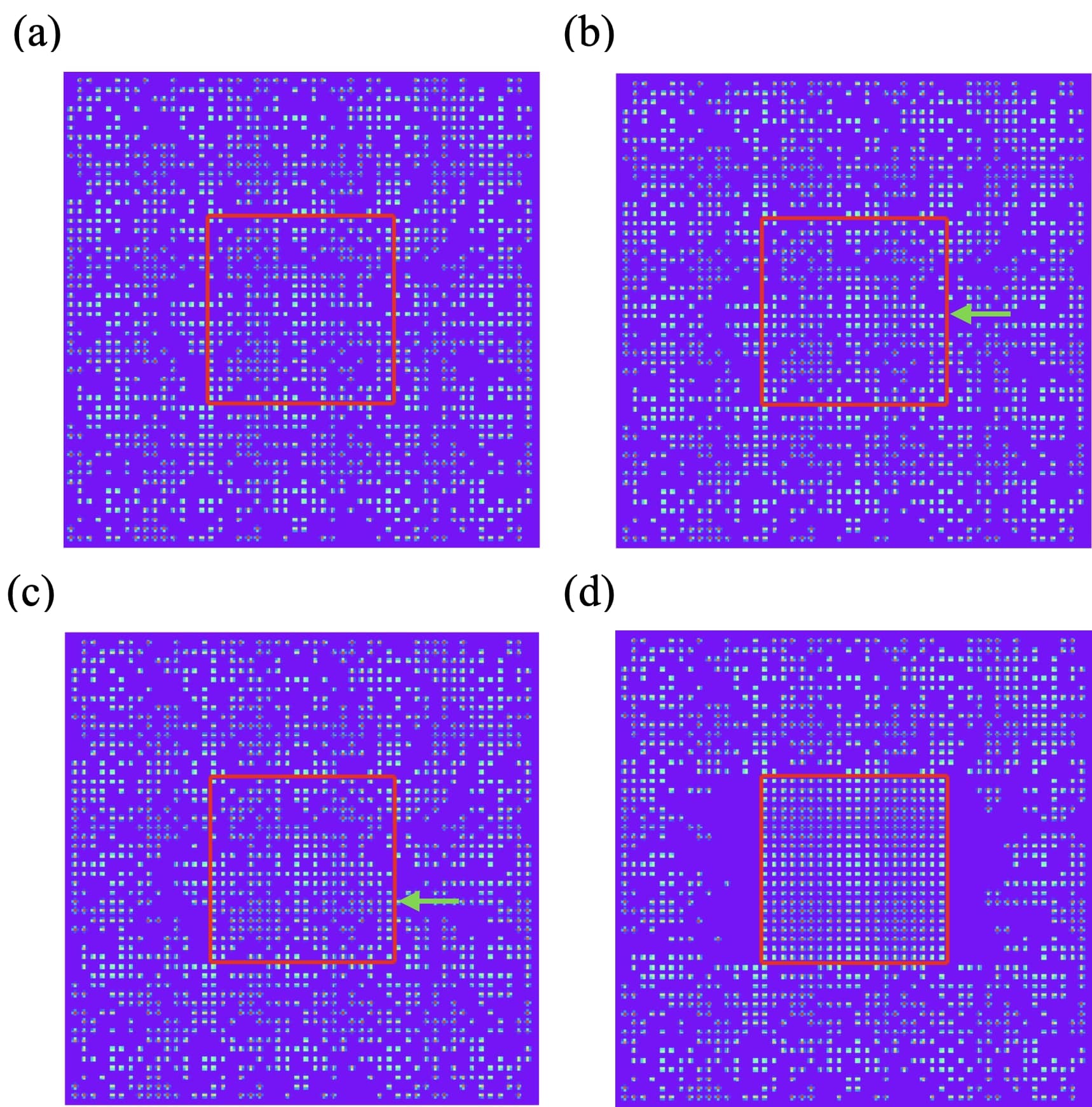}
    \caption{The visualization of the 2D tweezer rearrangement simulation. (a) shows the initial state of the 2D tweezer array. We can see the randomly positioned defect sites. (b) and (c) show the 2D array of tweezers rearranged by selecting by the number assigned to the row as the action value. (d) shows the termination state of the 2D tweezer array. A defect-free 2D array is formed in the central area.}
    \label{FIG:8}
\end{figure}

\begin{figure}[b]
    \centering
    \includegraphics[width=90mm]{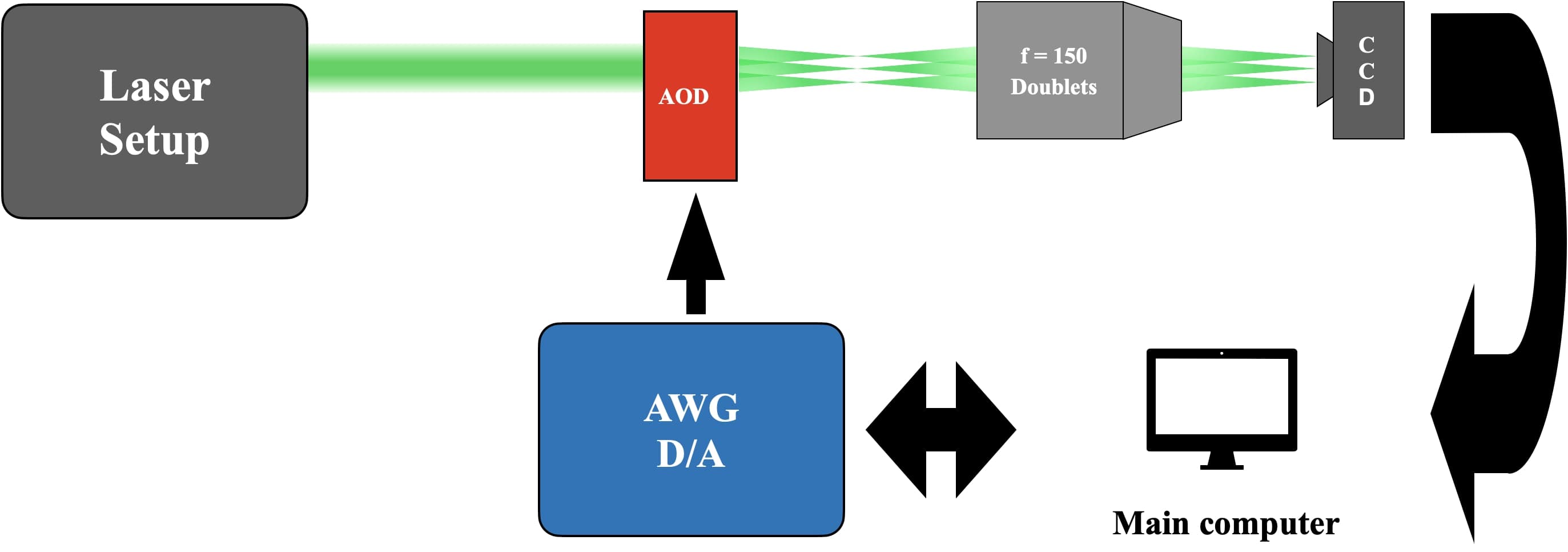}
    \caption{Schemetic of the actual physical setup for the tweezer rearrangement}
    \label{FIG:9}
\end{figure}
For the reinforcement learning, the state function is structured as a 2D binary array. Similar to the previous 1D array state configuration, each site with an atom is represented as 1, and a defect site is represented as 0, forming a 2D binary state composed of 1s and 0s. The action values are set as a discrete action space within the dimensions of the total rows and columns of the binary array. The state transition for action values is designed such that, for each time step, the corresponding row or column is selected sequentially in the 2D array, and the 1D tweezer rearrangement neural network operates until the selected 1D array reaches the Terminate state. The Terminate state is defined as a state where all the site within a central square area of 16\% of the total array are occupied with atoms. Reward values are assigned similarly to the 1D array rearrangement but only using negative reward, with a reward of $r=-0.1$ for each time step, aiming for the rearrangement to occur in the shortest time. If the state within the region of interest changes, a reward of $r=-0.01$ is assigned. Additionally, when the state transition persists and eventually reaches the Terminate state, the reward is set to 0. It is intended to create a situation where a defect-free square area, located centrally in the 2D array, using the 1D array neural network.

The neural network training algorithm for learning uses the same PPO algorithm as the 1D tweezer array. The CNN layer structure and fully connected layer structure for each network are also set the same, with the difference being the use of 2D CNN layers in this case. The state function \textit{s} is composed of the entire two-dimensional binary array representing atom occupancy in the tweezer array. The size of this function is expressed as $1\cdot(T_{\mathrm{size}})^2$, where $T_{\mathrm{size}}$ is the size of the 1D tweezer array. This state function, processed by the actor and critic networks' CNN filters, undergoes a reduction in dimensions as it passes through CNN layers with kernel sizes decreasing in sequence 7, 5, 3 and an increase in the number of filters 2, 4, 6. Batch normalization and ReLU further refine the data, and the final encoded data dimension is $1\cdot6\cdot(T_{\mathrm{size}}-12)^2$. The encoded state information, with a dimension of $1\cdot6\cdot(T_{\mathrm{size}}-12)^2$, is then flattened to become the input for the fully connected layers. For the actor network, the output is \emph{$\pi_{\theta}(s, a)$} with a dimension of $1\cdot2\cdot T_{\mathrm{size}}$, while for the critic network, the output is \emph{$V_{\phi}(s)$} with a dimension of $1\cdot1$. The neural network is updated similarly to the 1D tweezer array training, using the values output from the network.

The results of reinforcement learning for the 2D array are summarized in Figure \ref{FIG:7}. Figure \ref{FIG:7} depicts the scores and iterations for each episode of reinforcement learning for a total tweezer size of 50 by 50. As episodes progress, scores tend to converge to 0, and the number of iterations per episode decreases, indicating that training progresses smoothly. This confirms that pre-trained 1D tweezer neural networks can be implemented to train for 2D tweezer array rearrangement successfully. In Figure \ref{FIG:8}, the simulation results for a 50 by 50 2D tweezer array are visually presented.
\begin{figure*}[t]
    \centering
    \includegraphics[width=140mm]{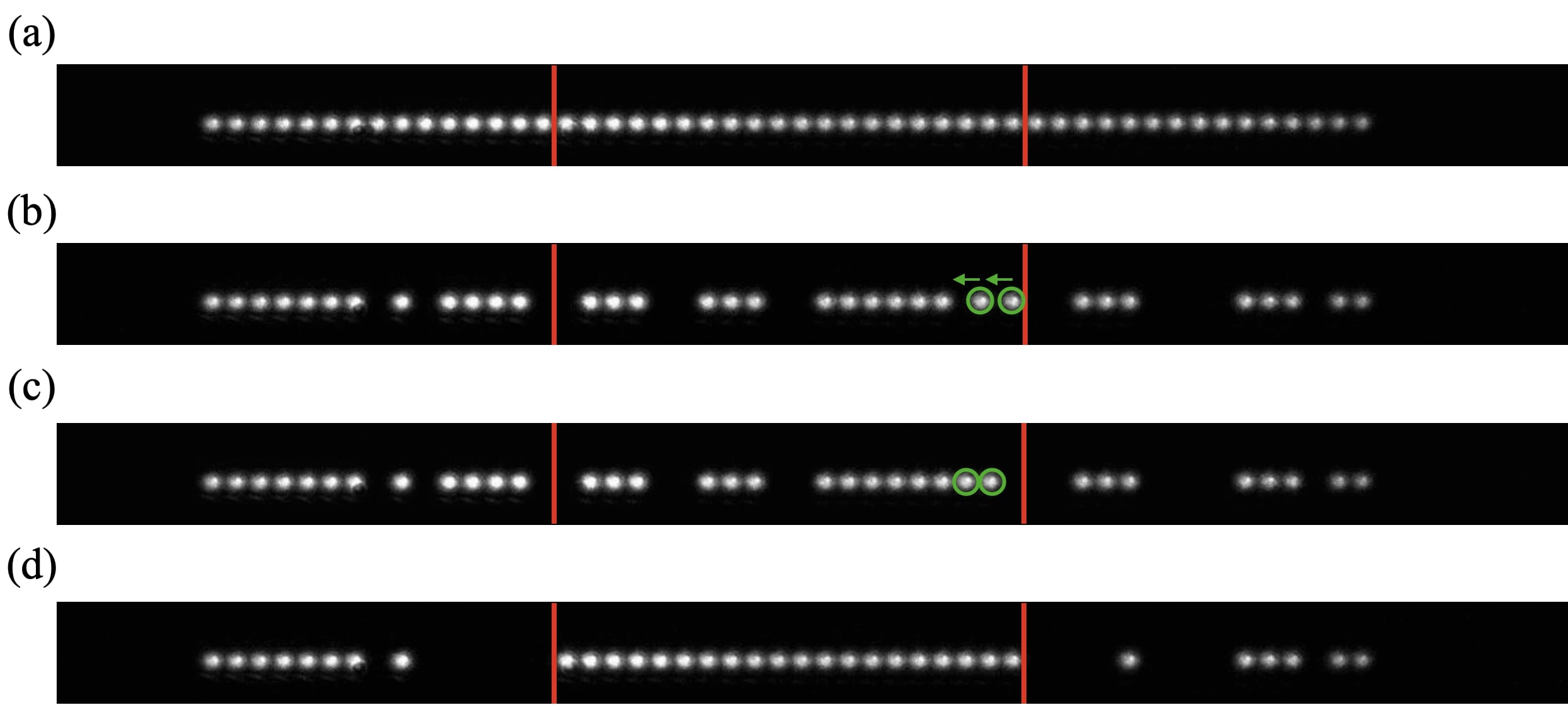}
    \caption{Demonstration of tweezer rearrangement in a physical setup. (a) the initial optical tweezer array before loading atoms. The size of the implemented optical tweezer array is 50. (b) the initial state after atoms are loaded into random tweezer sites. (c) the rearrangement of the tweezers using the trained neural network. When the state information is given to the neural network with action values of 33 and 34 (marked with green circles), the atoms located at the same index are moved sideways by one site. (d) the final array of the tweezers. 20 atoms, which is 40\% of the total number of tweezers, are aligned in a defect-free manner. The remaining atoms act as a reservoir, and when a region of interest, within the red vertical lines, becomes defective again, the neural network is reactivated and uses the atoms in the reservoir to fill in its position.}
    \label{FIG:10a}
    \label{FIG:10b}
    \label{FIG:10c}
    \label{FIG:10}
\end{figure*}
\section{Results of implementing neural networks in real-world physical environments}
The pre-trained neural network was implemented to the experimental setup to demonstrate the rearrangement of real optical tweezers. The trained neural network was ported using a PyTorch Just-In-Time compiler(JIT compiler), and the input information for the neural network was created by probabilistically capturing atoms in arbitrary sites, following the same method as in the simulation environment. The Arbitrary wave generator(AWG) card used to generate RF signals was Spectro instrument's M4i.6631-x8, and a single AOD, DTSX-400-532, was utilized. The main computer had an Intel i9-13900 CPU, an NVIDIA RTX-4080 GPU, and 64GB of RAM. The overall schematic of the physical environment can be seen in Figure \ref{FIG:9}. The configured binary array served as the input for the trained neural network, and for each state transition, the corresponding RF frequency to the selected tweezer is tuned to move the site. In Figure \ref{FIG:10}, the transition from the initial state with defect sites to the Terminate state is observed to be implemented in the actual optical setup. In the initial state(Figure \ref{FIG:10b}b), defects exist between each array, but when a state transition occurs, the neural network selects appropriate tweezers moves them. At each time step, the tweezers move one by one(Figure \ref{FIG:10b}b and \ref{FIG:10c}c), ultimately forming a defect-free state in the central 40\% of the total array size.

\section{Conclusion}
Reinforcement learning was introduced to address the optimization problem of optical tweezer array rearrangement, confirming successful navigation of the optimal path within the physical environment to reach the termination state. Our method involved treating the entire optical tweezer array as a single agent, whereas in reality, considering each trapped atom in the optical tweezer array as an individual agent in a multi-agent environment would be more effective \cite{RN57,RN58,RN59}. This multi-agent approach could lead to moving more atoms within a single time step. However, the stochastic and dynamic nature of the atomic array requires changing the number of agents in every episodes for the multi-agent framework, which is challenging in reinforcement learning \cite{RN61}. In this experiment, we demonstrated that simplifying the optical tweezer rearrangement problem to a single agent allows for a reduction in the number of neural network parameters while still effectively solving the problem using reinforcement learning. The trained neural network was successfully transferred to the device, enabling the implementation of optical tweezer rearrangement in the actual physical setup. This work presents the potential of reinforcement learning to meet various technical requirements in experiments in the field of AMO physics.

\section*{Declaration of competing interest}
There are no conflicts of interest to declare.

\section*{Declaration of generative AI and AI-assisted technologies in the writing process}
During the preparation of this work the authors used OpenAI ChatGPT in order to translate. After using this service, the authors reviewed and edited the content as needed and take full responsibility for the content of the publication.

\section*{Acknowledgments}
Authors apprecite Q-han Park for insightful discussions about the project. Special appreciation is given to Kikyeong Kwon, Seunghwan Roh, Giseok Lee, and Youngju Cho, whose expert advice on experimental design and statistical analysis was indispensable in shaping the research methodology and interpreting the results. We are also grateful to Hyunjun Jang and Donkyu Lim for providing their extensive knowledge and technical support in the field of artificial intelligence. The authors acknowledge support from the National Research Foundation of Korea under grant numbers \seqsplit{2020R1A4A1018015}, \seqsplit{2021M3H3A1085299}, \seqsplit{2022M3E4A1077340}, \seqsplit{2022M3C1C8097622}, \seqsplit{2022M3H3A106307411}. 

\bibliographystyle{elsarticle-num.bst}
\bibliography{refs.bib}
\biboptions{sort&compress}
\end{document}